\documentclass[USenglish,10pt]{article}
\usepackage[utf8]{inputenc}%(only for the pdftex engine)
\usepackage[small]{dgruyter}
\usepackage{microtype}
\usepackage{subfig}
\usepackage[colorinlistoftodos]{todonotes}
\begin{document}
  \articletype{Research Article}
  \author*[1,2,3]{Raghu Dharmavarapu}
  \author[4]{Ken-ichi Izumi}
  \author[4]{Ikufumi Katayama}
  \author[2,3]{Soon Hock Ng}
  \author[5]{Jitraporn Vongsvivut}
  \author[5]{Mark J. Tobin}
   \author[6,7]{Aleksandr Kuchmizhak}
  \author[8,9]{Yoshiaki Nishijima}
  \author*[1]{Shanti Bhattacharya}
  \author*[2,3,9,10]{Saulius Juodkazis}
  \runningauthor{Raghu Dharmavarapu et.al}
  \affil[1]{Centre for NEMS and Nanophotonics (CNNP), Department of Electrical Engineering, Indian Institute of Technology Madras, Chennai 600036, India}
  \affil[2]{Centre for Micro-Photonics, Faculty of Science, Engineering and Technology, Swinburne University of Technology, Hawthorn VIC 3122, Australia}
  \affil[3]{Melbourne Centre for Nanofabrication, ANFF, 151 Wellington Road, Clayton, VIC 3168, Australia}
  \affil[4]{Physics Department, Yokohama National University, 79--5 Tokiwadai, Hodogaya-ku, Yokohama 240--8501, Japan}
  \affil[5]{Infrared Microspectroscopy Beamline, Australian Synchrotron, Clayton, Victoria 3168, Australia}
  \affil[6]{Institute of Automation and Control Processes, Far Eastern Branch, Russian Academy of Sciences, Vladivostok 690041, Russia}
 \affil[7]{Far Eastern Federal University, Vladivostok 690090, Russia}
  \affil[8]{Department of Electrical and Computer Engineering, Graduate School of Engineering, Yokohama National University, 79--5 Tokiwadai, Hodogaya--ku, Yokohama 240--8501, Japan}
\affil[9]{Institute of Advanced Sciences, Yokohama National University, 79-5 Tokiwadai, Hodogaya-ku, Yokohama 240-8501, Japan.}
\affil[10]{Tokyo Tech World Research Hub Initiative (WRHI), School of Materials and Chemical Technology, Tokyo Institute of Technology, 2-12-1, Ookayama, Meguro-ku, Tokyo 152-8550, Japan}

  \title{Dielectric cross-shaped resonator based metasurface for vortex beam generation in Mid-IR and THz wavelengths}
  \runningtitle{Nanophotonics}
  \subtitle{}
  \abstract{Metasurfaces are engineered thin surfaces comprising two dimensional (2D) arrays of sub-wavelength spaced and sub-wavelength sized resonators. Metasurfaces can locally manipulate the amplitude, phase and polarization of light with high spatial resolution.  In this study, we report numerical and experimental results of a vortex-beam-generating metasurface fabricated  specifically for infrared (IR) and terahertz (THz) wavelengths. The designed metasurface consisted of a 2D array of dielectric cross-shaped resonators with spatially varying length, thereby providing desired spatially varying phase shift to the incident light. The metasurface was found to be insensitive to polarization of incident light. The dimensions of the cross-resonators were calculated using rigorous finite difference time domain (FDTD) analysis. The spectral scalability  via physical scaling of meta resonators was demonstrated using two vortex generating optical elements operating at 8.8~$\mu$m (IR) and  0.78~THz (Terahertz). The vortex beam generated in the mid-IR spectral range was imaged using FTIR imaging miscroscope  equipped with a focal plane array (FPA) detector. This design could be used for efficient wavefront shaping as well as various optical imaging applications in mid-IR spectral range, where polarization insensitivity is desired.}
  \communicated{}
  \dedication{}
  \received{10/04/2019}
  \accepted{02/06/19}
  %\journalname{Nanophotonics (Accepted on 2 Jun 2019)}
  \journalyear{2019}
  \journalvolume{xx}
  \journalissue{xx}
  \startpage{1}
  \aop
  \DOI{xxx.xxx}

\maketitle
\section{Introduction}

Optical elements to be used in the IR and THz spectral ranges with wavelengths spanning from tens-of-$\mu$m to sub-mm, and polarisation optical elements based on refraction become non-practical due to the shortage of highly transparent materials in these wavelength regimes. In addition, fabrication can often be challenging for high aspect-ratio patterns over large macroscopic areas. IR radiation and THz rays have been used and applied extensively in the optical research and development community in the past decades. While THz technology has gained an increasing demand for applications in sensing and non-invasive imaging fields~\cite{jansen2010terahertz,redo2008terahertz,bitzer2008terahertz}, the IR-based techniques have been established as a molecular characterisation tool for a wide range of applications in microstructural analysis, spectroscopy and imaging~\cite{honda2019paracetamol,honda2018simple}. 

%_______________________Fig. 2
% \begin{figure*}[tb]
% \centering
% \subfloat[]{
% \includegraphics[width = 2.in]{IRPhCurve.pdf}
% \label{fig:2a}
% }
% \subfloat[]{
% \includegraphics[width = 2.in]{THzPhCurve.pdf}
% \label{fig:2b}
% }
% \caption{Look up tables for IR and THz cross resonators: (a) The length $L$ of the cross arm is varied from 1~$\mu$m to 6.5~$\mu$m to achieve 0-2$\pi$ transmission for IR design (b) The length $L$ of the cross arm is varied from 180~$\mu$m to 280~$\mu$m to achieve 0--2$\pi$ transmission for the THz design. Inset: Schematic of the vortex generator}
% \label{fig:2}
% \end{figure*}

Several methods were proposed to manufacture basic optical elements for these wavelengths, such as refractive and diffractive lenses, polarizers and beam splitters~\cite{tonouchi2007cutting,mittleman2013sensing}. Recently, 3D printing was employed to manufacture non-conventional diffractive optical elements to generate complex light beams such as Bessel, Airy and vortex beams in THz and IR wavelengths~\cite{wei2015generation,liu20163d,machado2019multiplexing}. Although 3D printing technique is often considered to be a good solution for generating custom light beams, this approach is still limited for applications that require a higher fabrication throughput due to a direct write character. There is considerably less work on THz and IR complex light generation despite their huge potential applications. Diffractive optical elements (DOEs) could be considered as a solution because they reduce the volume that should be 3D printed. However, the efficiency of DOEs is significantly lower, leading to a major limitation especially for applications in the THz spectral window.     

Laguerre Gaussian beams also known as vortex beams are interesting complex beams in optics, which possess
helical wavefronts and an on-axis phase singularity~\cite{nye1974dislocations,barnett1994orbital}. In addition, composite vortex beams possess a self-healing property within the Rayleigh range~\cite{srinivas2018investigation}. Owing to these interesting properties, vortex beams have been used in many applications, such as optical manipulation of micro particles~\cite{he1995direct,ng2010theory}, fibre optical communication~\cite{willner2015optical}, STED microscopy~\cite{willig2006sted}. Currently,  the most popular techniques to generate these beams, in the visible wavelength, include computer generated holograms~\cite{heckenberg1992generation}, spatial light modulators~\cite{matsumoto2008generation}, Spiral
Phase Plates (SPP)~\cite{beijersbergen1994helical}, and so on. In 2012, P. Genevet et al. employed a V-shaped antenna-based metasurface~\cite{genevet2012ultra} to build a vortex generating  SPP  in the IR spectral region. A year later, the same design concept was extended to the THz region and used to fabricate vortex generators~\cite{hu2013ultrathin} and a modified version of antenna was proposed to generate vortex beams in THz region~\cite{he2013generation}. Geometric metasurfaces, which work on the pancharatnam-Berry phase phenomenon are used to generate both scalar~\cite{devlin2017arbitrary} and vector~\cite{yue2016vector} vortex beams. Recently, polarization-sensitive generation and modulation~\cite{yan2019generation} of OAM beams using dielectric metasurfaces based on silicon fin structures. Metasurfaces have recently become a platform to realize and apply optical spin-hall effect\cite{xiao2016spin}. Metasurfaces can be built using several type of meta-atoms, such as V-shaped antennas~\cite{kildishev2013planar}, cylindrical disks~\cite{chong2015polarization} or nano-fin structures~\cite{khorasaninejad2016metalenses}. 
\par
In this study, we develop a vortex beam generator for the mid-IR spectral range ($\sim 8.8~\mu$m), using cross-shaped resonators~\cite{dharmavarapu2018all}, which can support both Mie-type electric and magnetic dipole resonances to realize phase manipulation and to completely suppress reflection losses. The idea behind choosing this wavelength is to use the generated vortex beam to measure the absorbance of secondary protein structures such as beta sheets in silk which also lie at in the same wavelength region~\cite{17m356,17mre115028,17sr7419,19bjn922}. The generated vortex beam was captured, for the first time, using FTIR imaging microscope with a focal plane array (FPA) detector. To show the versatility and scalability of this design, we scaled the IR design and generated a vortex beam of charge 1 at 0.78~THz. Transmission response of both s- and p-polarizations showed a polarization insensitivity for this design. Finite difference time domain (FDTD) simulation, fabrication, and optical characterisation results are presented to support the experimental results. 

\section{Results and discussion}

\subsection{Design of the unit cell}

\begin{figure*}[!h]
\centering
\subfloat[]{
\label{fig:1b}
\includegraphics[width = 4.5in]{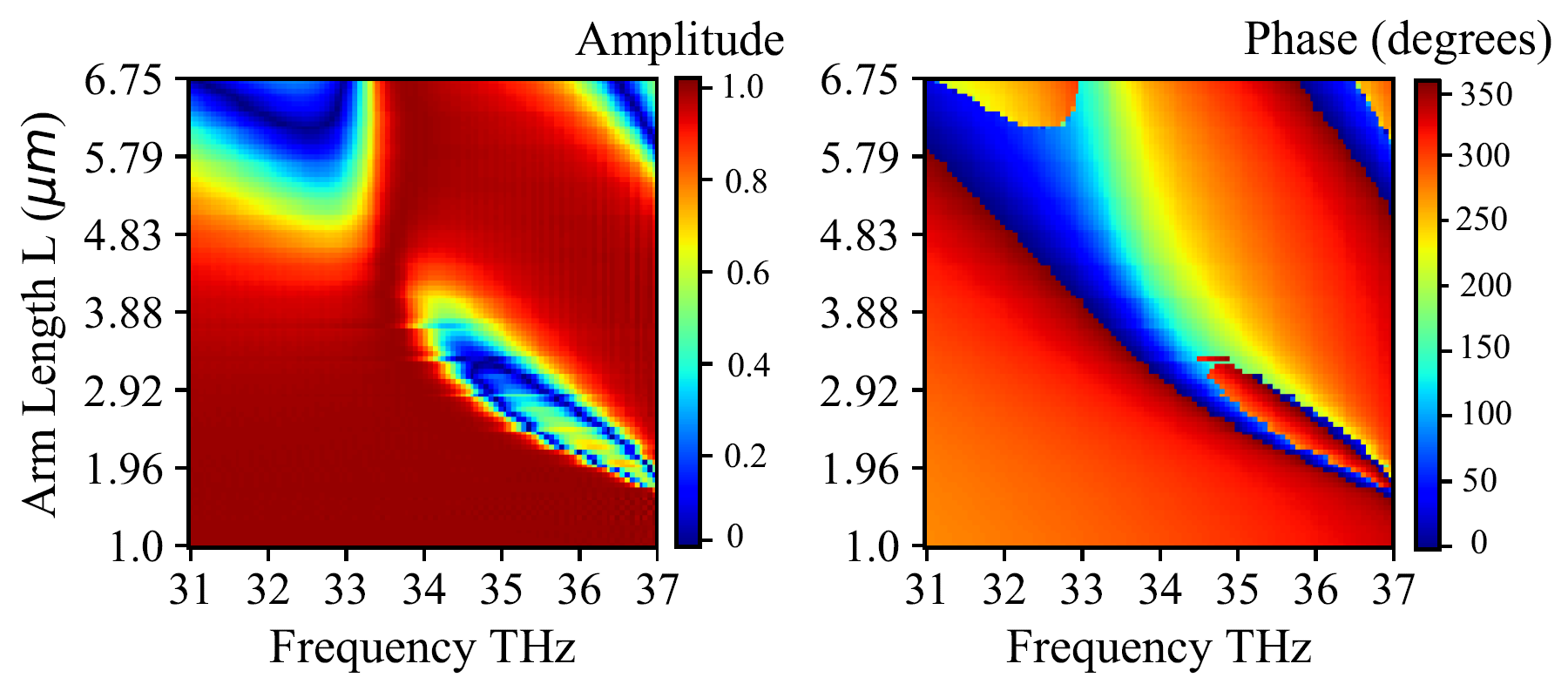}
}
\\
\subfloat[]{
\includegraphics[width = 1.75in]{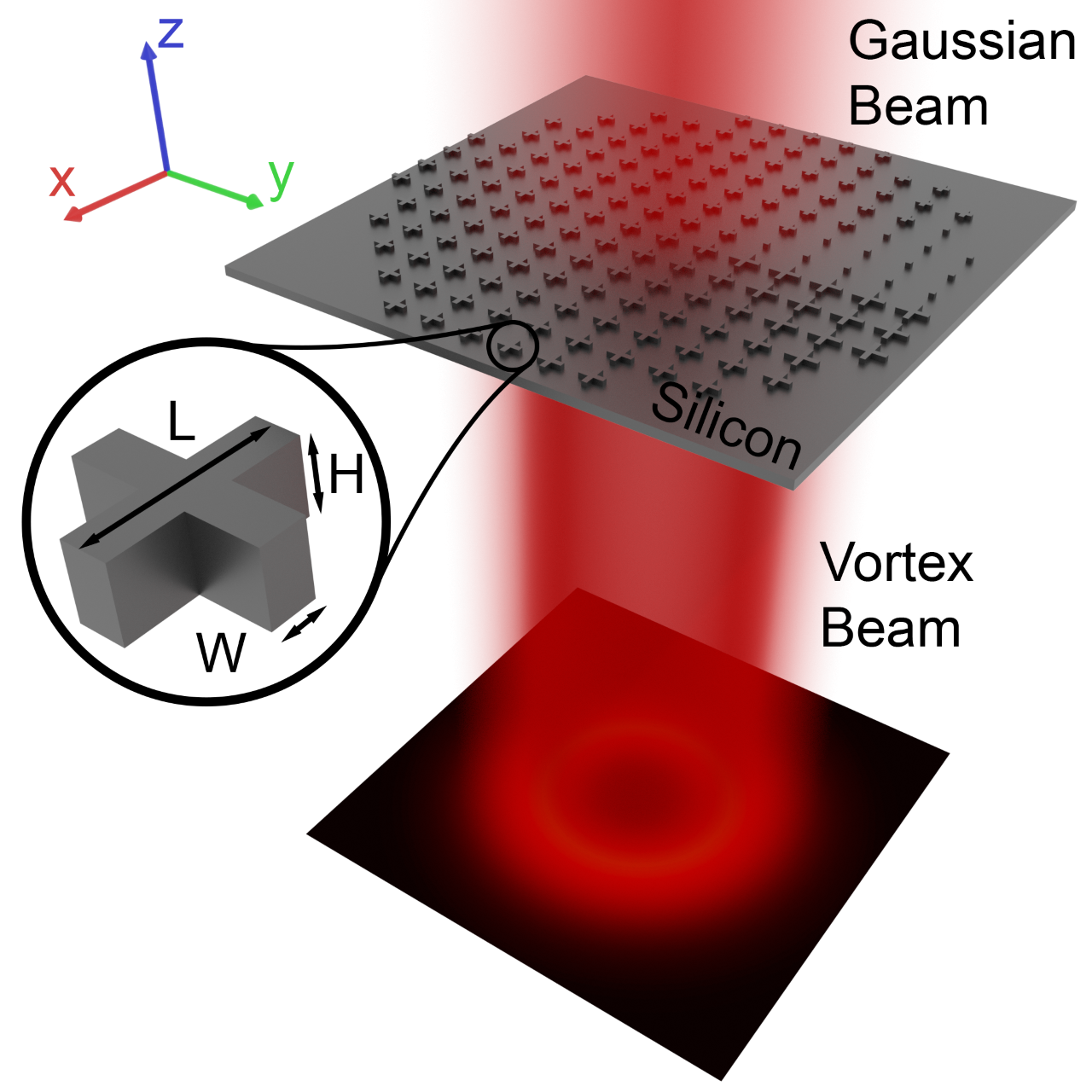}
\label{fig:1c}
}
\subfloat[]{
\includegraphics[width = 2.5in]{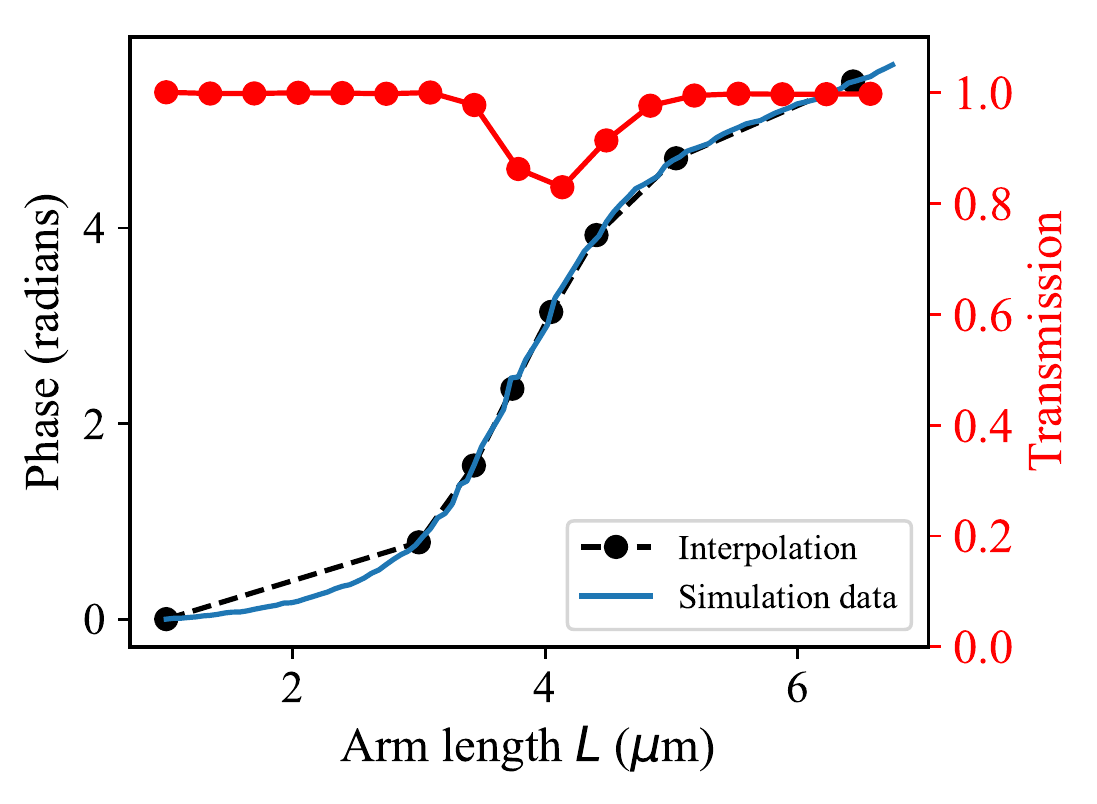}
\label{fig:1d}
}
\caption{
\label{fig:1}
(a) Simulated transmission amplitudes and corresponding transmission phases as a function of arm length, $L$, over the frequency range 31-37~THz. Spectral position near $\sim 34$~THz was selected for design of vortex generator due to the constant transmission amplitude and possibility of $2\pi$ phase control. (b) Schematic of the vortex generator layout, silicon cross meta-atom in the inset (c) Look up table for IR cross resonators: The length $L$ of the cross arm is varied from 1~$\mu$m to 6.5~$\mu$m to achieve 0-2$\pi$ transmission for IR design}
\end{figure*}

In this section, we discuss the design parameters and dimensions of silicon cross resonators to achieve high transmission and 0 to 2$\pi$ phase response for full wavefront control. A schematic of the silicon cross resonator is shown in the inset of Fig.~\ref{fig:1c}. Silicon was chosen due to its high relative permittivity of $\epsilon_r = 13.5$ and its low loss nature in the THz  regime. The design frequency is chosen to be 34~THz, which lies in one of the absorption bands of silk in the IR spectral range. The following cross resonator dimensions are optimized to get maximum transmission and $2\pi$ phase coverage at the chosen operational frequency. The $2\pi$ phase coverage is achieved by changing the length of the resonator from 1.25~$\mu$m to 6.5~$\mu$m. The width, height and lattice constant of the unit cell are fixed at 1$~\mu$m, 1.85~$\mu$m and 7.25~$\mu$m respectively.

%_________________________Fig. 1

We considered the electromagnetic response of one resonator with periodic boundary condition as the spacing between the resonators is sub-wavelength, which allows very low coupling between adjacent resonators. Each resonator can support electric and magnetic resonances, when the dimensions become comparable to the wavelength of the incident light~\cite{evlyukhin2010optical}. High transmission can be achieved by tuning the dimensions such that the two resonances overlap spectrally and cancel out the back scattered light. In our simulations, the refractive index model of silicon is fitted to Palik data and the incident wave is assumed to be a plane wave propagating along the $z$ axis with electric field along the $x$ direction. The transmission amplitude and phase is extracted from the S21 parameter of the FDTD simulation. The FDTD-extracted electric field amplitude and phase data, as a function of varying cross length (1.25~$\mu$m-6.5~$\mu$m) and operating frequency(31-37~THz), are shown in Fig.~\ref{fig:1b}. The plot in Fig.~\ref{fig:1d} shows the variation of transmission amplitude and phase with respect to changing cross length from 1.25~$\mu$m to 6.5~$\mu$m, at the desired frequency 34~THz ($\lambda$ = 8.8~$\mu$m). 

For the THz vortex generator, we chose 0.78~THz as the operating frequency. This corresponds to a wavelength of 410~$\mu$m. This wavelength was selected as the band close to the coherent synchrotron radiation (CSR) at 0.3-0.4~THz. CSR is an unique highest brightness radiation source at the IR-beamline at the Australian synchrotron which can benefit from set of optical elements to control polarization, intensity, and wavefront. We used linear scaling approach to find the desired starting dimensions for the THz metasurface. The wavelength scaling factor is 46.5, which is calculated by taking the ratio of the two wavelengths. We found the maximum of length for the cross resonator to be approximately 300~$\mu$m  by applying the same scaling factor to the length of the cross resonator. By using further FDTD optimization, we narrowed down the desired length range to 180~$\mu$m-275~$\mu$m to achieve 0 to 2$\pi$ phase coverage at 0.78~THz. The width, height and lattice constant of the cross were fixed at 70$~\mu$m, 150~$\mu$m and 380~$\mu$m respectively. 

\subsection{Mid-IR vortex generator}

A metasurface for vortex beam generation at mid IR wavelengths is constructed as follows. The phase function of the vortex generator is given by a helical function as,
\begin{equation}
\Phi(x,y) =l \tan^{-1}(y/x),
\label{eq:1}
\end{equation}
where ($x,y$) are the coordinates in the metasurface plane and $l$ is the topological charge of the vortex beam. The continuous phase profile of the vortex generator given in the Eq.~\ref{eq:1} is discretized to eight phase levels and eight cross dimensions  with nearly equal phase steps of $\pi/4$ to cover the full $2\pi$ range are selected from the look-up table plot Fig.~\ref{fig:1d}. It can be seen that all the resonators have high transmission with a mean variation of only 5\%. The schematic of the vortex generator is shown in Fig.~\ref{fig:1c}.  

%___________________________Fig. 2
\begin{figure*}[!ht]
    \centering
    \subfloat[]{
    \includegraphics[width = 2.5in]{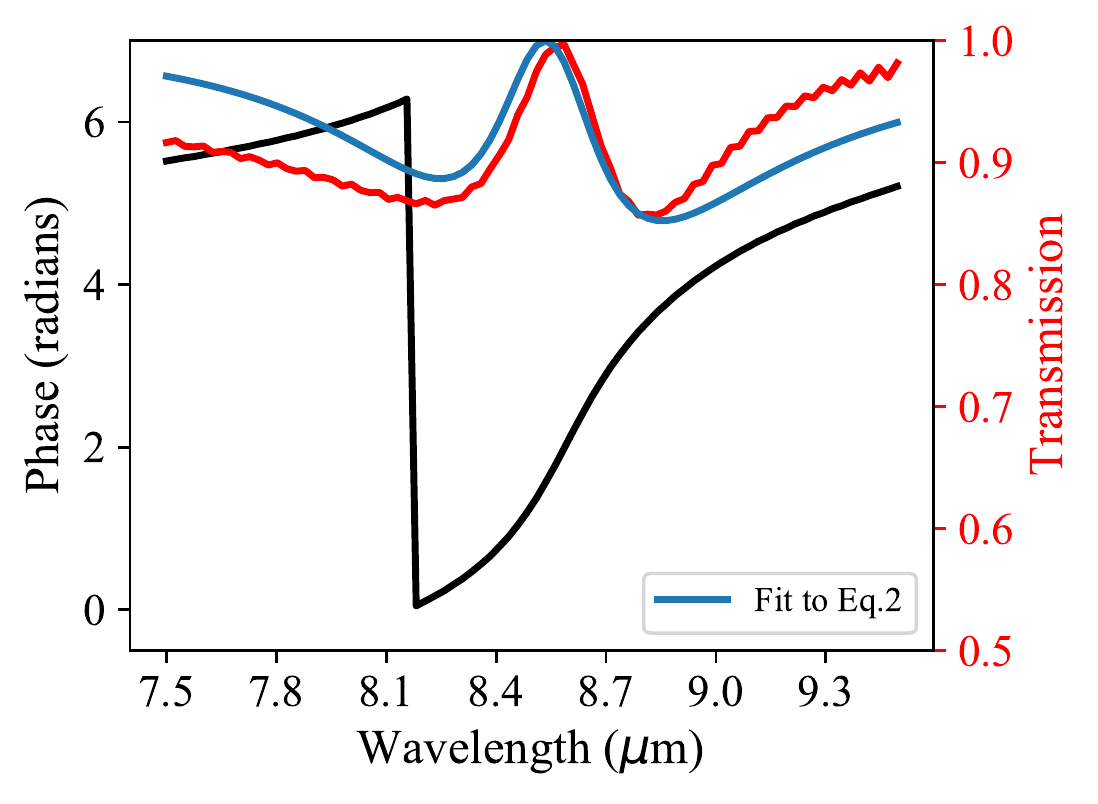}
    \label{fig:tatp}
    }
    \subfloat[]{
    \includegraphics[width = 2.5in]{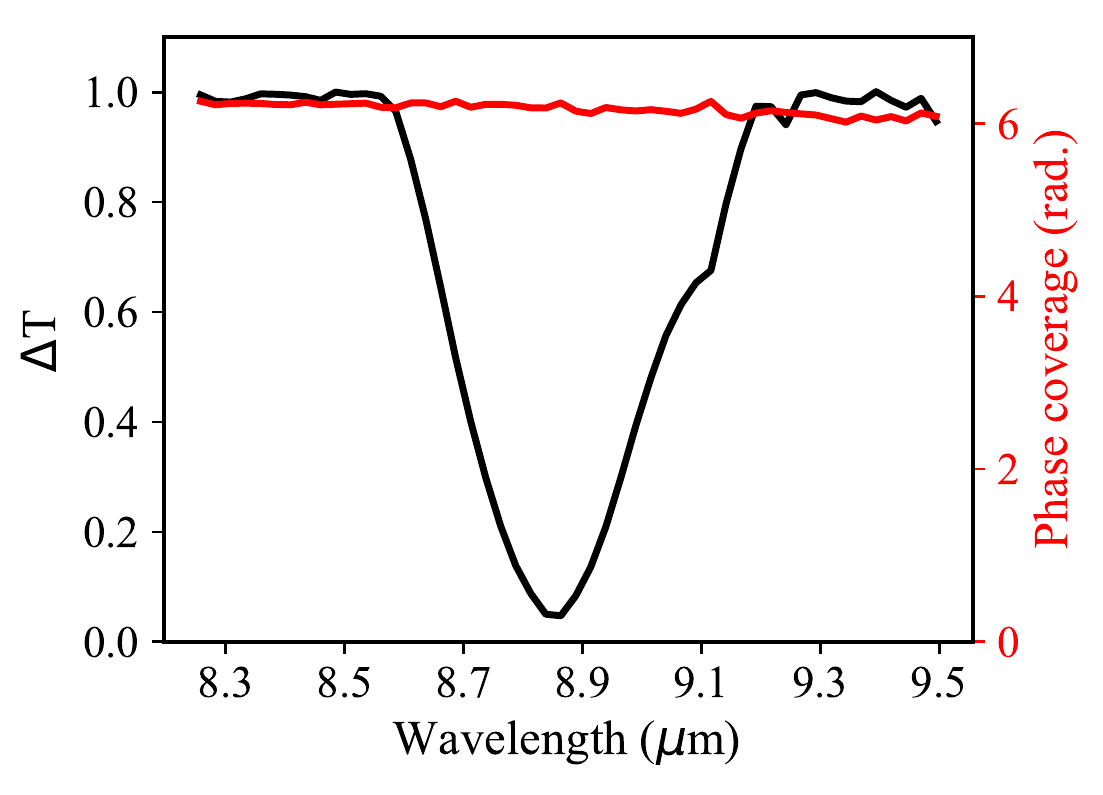}
    \label{fig:layout}
    }
    \caption{ (a) FDTD simulated  amplitude and phase for a cross resonator with arm length $L = $3$\mu$m; a curve fitting to Eq~\ref{eq:2} is plotted by blue line. (b) Maximum change in the transmission $\Delta T$ and phase coverage among the eight chosen cross resonators. }
\end{figure*}
To gain a intuitive insight into the phase tuning phenomenon of the cross resonator, a simple analytical model~\cite{decker2015high} is presented here. Each of the cross resonators can be considered as a coupled electric and magnetic dipole excited by an incident electromagnetic wave. The far-field electric field of the metasurface will be the sum of incident field and the electric and magnetic dipole radiation coming from these individual cross resonators. Therefore, the transmission coefficient of each resonator can be given by
\begin{equation}
    t(\omega) = 1+\frac{2j\gamma_e\omega_e}{\omega_e^2-\omega^2-2j\gamma_e\omega_e}
    +\frac{2j\gamma_m\omega_m}{\omega_m^2-\omega^2-2j\gamma_m\omega_m},
    \label{eq:2}
\end{equation}
where $\omega$ is the frequency of incident beam, $\omega_e$ and $\omega_m$ are
the electric and magnetic resonance frequencies, and $\gamma_e$ and $\gamma_m$ are the damping factors of the electric and magnetic dipoles.
Fig.~\ref{fig:tatp} shows the transmission phase and amplitude simulated using FDTD(red curve) and curve fit to the analytical model given in Eq.~\ref{eq:2} for the cross resonator with an arm length of 3~$\mu$m. For this cross resonator, we have extracted the electric and magnetic resonance frequencies $\omega_e,\omega_m$ and the damping factors $\gamma_e, \gamma_m$ to be 35.0~THz, 35.1~THz and 2.13, 0.72 respectively using curve fitting. The difference between the analytical model and FDTD data can be attributed to the finite mesh settings of the FDTD solver. Fig.~\ref{fig:layout} shows the maximum variation in transmission amplitude ($\Delta$T) and phase coverage ($\Delta\Phi$) among the eight chosen cross resonators over the wavelength range (8.3-9.5~$\mu$m). Even though the resonators are almost providing the desired 2$\pi$ (gradually decreasing) phase coverage over the entire frequency band, all the cross resonators at a specific frequency do not have the same transmission, which can be clearly seen from the dip in Fig.~\ref{fig:layout}. If we consider that a ~$20\%$ variation in transmission among the resonators is acceptable, we can use this design over a bandwidth of  $\approx$170~nm, around 8.85~$\mu$m.

%__________________Fig. 5
\begin{figure*}[tb]
\centering
\subfloat[]{
\includegraphics[height = 1.5in]{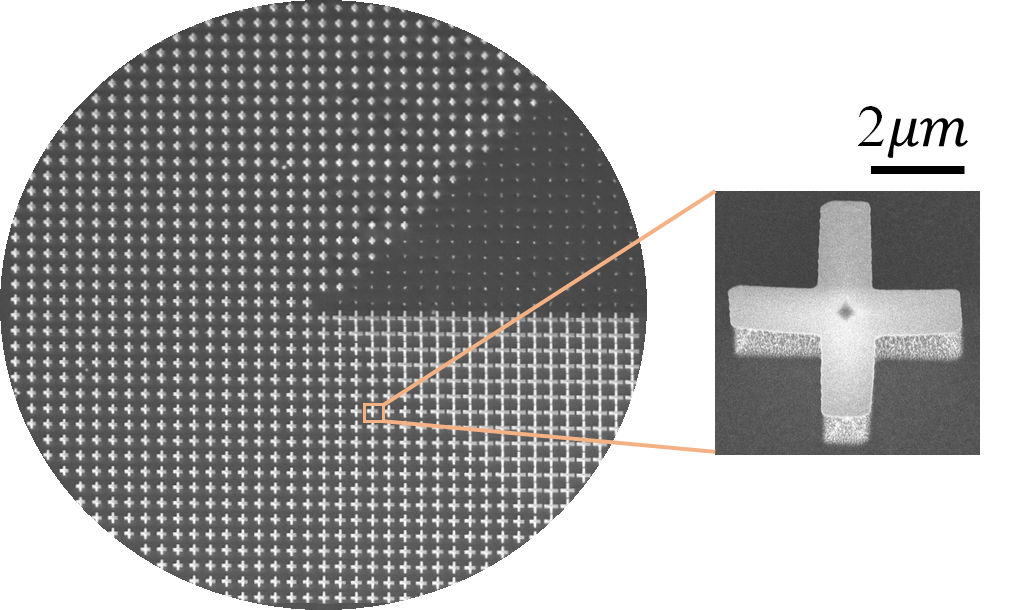}
\label{fig:9a}
}
\subfloat[]{
\includegraphics[height = 1.5in]{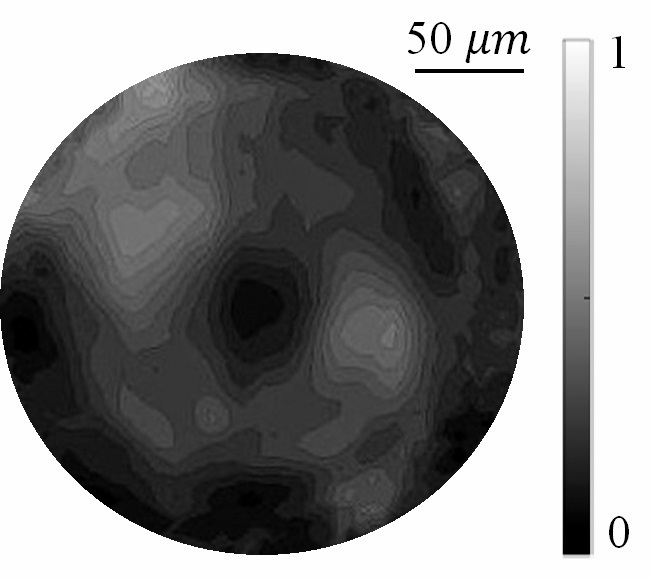}
\label{fig:9b}
}
\caption{(a) SEM image of meta SPP for $l = +1$ topological charge; the inset shows the single building block. (b) Vortex beam image captured on the $64\times 64$~pixel FPA detector.}
\label{fig:9}
\end{figure*}
We used a 500~$\mu$m thick silicon wafer to fabricate the IR vortex generator. The device was fabricated  using electron beam lithography (Vistec EBPG) followed by anisotropic dry etching using standard Bosch process, with photoresist acting as the etch mask. The etching recipe was optimized to get the desired cross height of 1.85~$\mu$m. 
A SEM image of the fabricated vortex plate is shown in
Fig.~\ref{fig:9a}, where the inset was captured at higher magnification to show a single cross meta-atom. 
\subsubsection{FPA-FTIR imaging of the vortex generator}

The IR images of the vortex beam were acquired with an offline FPA-FTIR microspectroscopic instrument at the Australian Synchrotron Infrared Microspectroscopy (IRM) Beamline. The system consisted of a Bruker Hyperion 3000 FTIR microscope (Bruker Optik GmbH, Ettlingen, Germany), equipped with a liquid-N2 cooled $64\times 64$~element FPA detector and a matching $15\times$~objective and condenser (NA = 0.40), which was coupled to a Vertex 70 FTIR spectrometer (Bruker Optik GmbH, Ettlingen, Germany) containing an internal thermal (Globar$^{TM}$) IR source.

FPA-FTIR images were collected in transmission mode within a 4000 to 800 cm$-1$ spectral region as a single FTIR image covering a sampling area of $180\times 180$ ~$\mu$m$^2$. Each FTIR spectral image consisted of a $64\times 64$~array of spectra, resulting from each square of the detector on the $64\times 64$~element FPA array. As a consequence, a single spectrum contained in a FTIR image represented approximately $2.8\times 2.8$ ~$\mu$m$^2$ area on the sample plane.

For each image, high-quality FTIR spectral images were collected at 4 cm$^{-1}$ resolution, with 128 co-added scans, Blackman-Harris 3-Term apodization, Power-Spectrum phase correction, and a zero-filling factor of 2 using OPUS 7.2 imaging software (Bruker). Background measurements were taken using the same acquisition parameters prior to sample spectral images, by focusing on a clean surface area of the substrate without the vortex pattern. All spectra were analyzed using OPUS v7.2 software, by integrating the area under the narrow spectral range around 1136~cm$^{-1}$($\lambda = 8.8~\mu$m). Figure~\ref{fig:9b} shows the vortex beam captured on the FPA detector. The low image quality can be attributed to the usage of only 8 steps in the SPP and lower resolution of the FPA sensor itself. A higher number of phase steps would improve the contrast of vortex intensity profile.

\subsection{THz vortex generator}

\begin{figure*}[!h]
    \centering
    \subfloat[]{
    \includegraphics[height = 3.75cm]{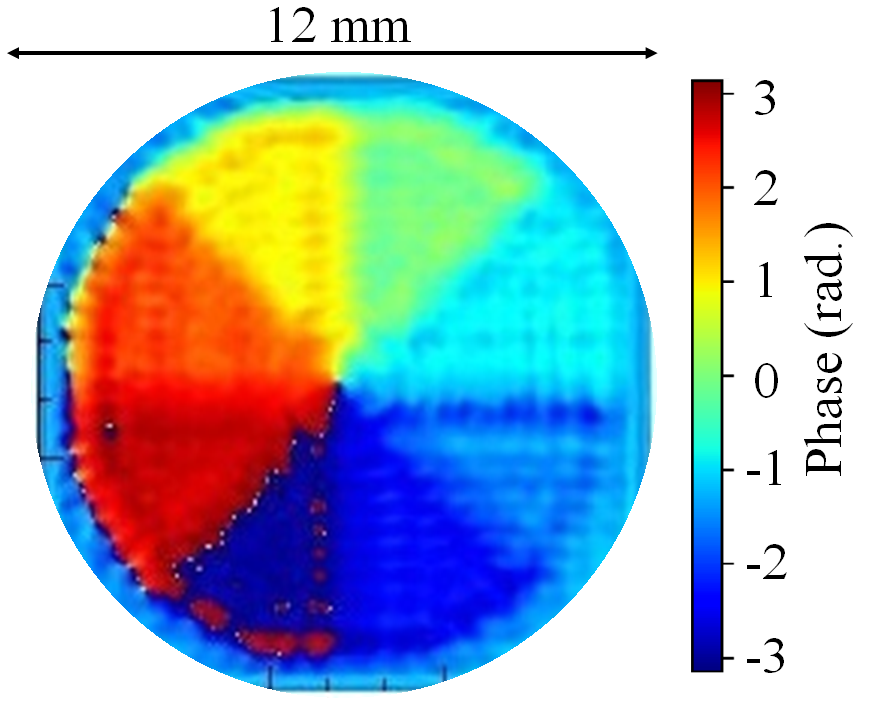}
    \label{fig:4a}
    }
    \hspace{1cm}
    \subfloat[]{
    \includegraphics[height = 3.5cm]{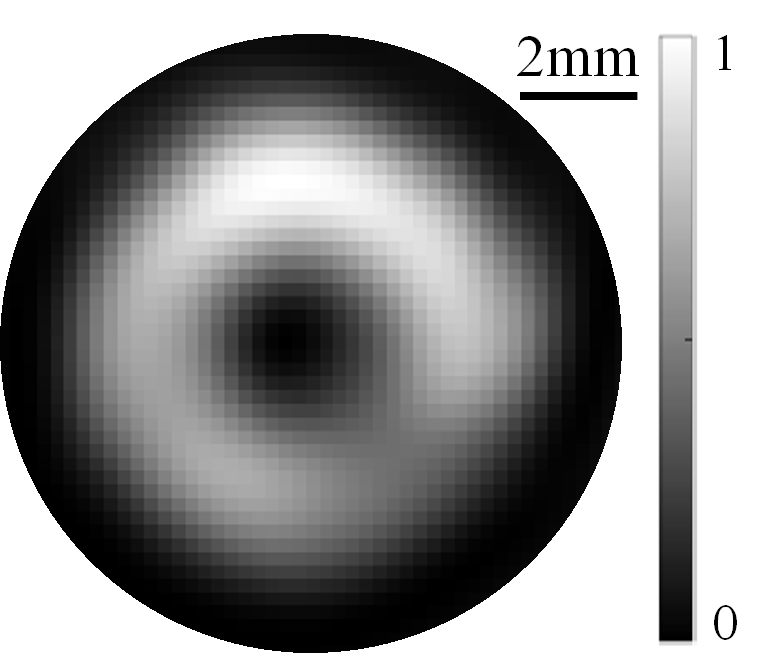}
    \label{fig:4b}
    }
    \caption{Azimuthal phase variation from 0 to 2$\pi$(a) and electric field intensity $|E|^2$ (b) distribution of the THz vortex beam generator; simulation frequency is 0.78~THz.}
    \label{fig:4}
\end{figure*}

We designed a SPP with a diameter 1.2~cm consisting of eight phase levels in its eight octants. Each of the octant contains an array of cross resonators such that the phase varies from 0 to 2$\pi$ azimuthally across vortex generator. Such an optical element has dimensions useful for a range of applications in THz experiments. The FDTD simulation of the  transmitted phase and electric field amplitude are shown in Fig.~\ref{fig:4a} and \ref{fig:4b} respectively. The designed spiral phase was found to indeed follow the required $2\pi$ azimuthal span. We experimentally show the high transmission and polarization independence for this metasurface. However, we could not image the vortex generator in a raster scan mode as our experimental setup is not capable of such mapping.

To simulate the SPP and the doughnut intensity pattern of the vortex generator, we created its GDSII layout(Schematic shown in Fig.~\ref{fig:1c}) using in-house developed software MetaOptics~\cite{raghu}. MetaOptics uses FDTD transmission phase vs varying dimension data of a meta-atom and creates GDSII layouts of any phase distribution by placing a meta-atom that gives the desired transmission phase at that pixel of the phase mask. The generated GDSII file is loaded in the Lumerical FDTD solver for simulation. The corresponding simulated phase and doughnut intensity pattern are shown in Fig.~\ref{fig:4a} and \ref{fig:4b}.

%___________________________Fig. 6

%________________________________Fig. 7
\begin{figure*}[tb]
\centering
\subfloat[]{
\label{fig:5a}
\includegraphics[height = 3.5cm]{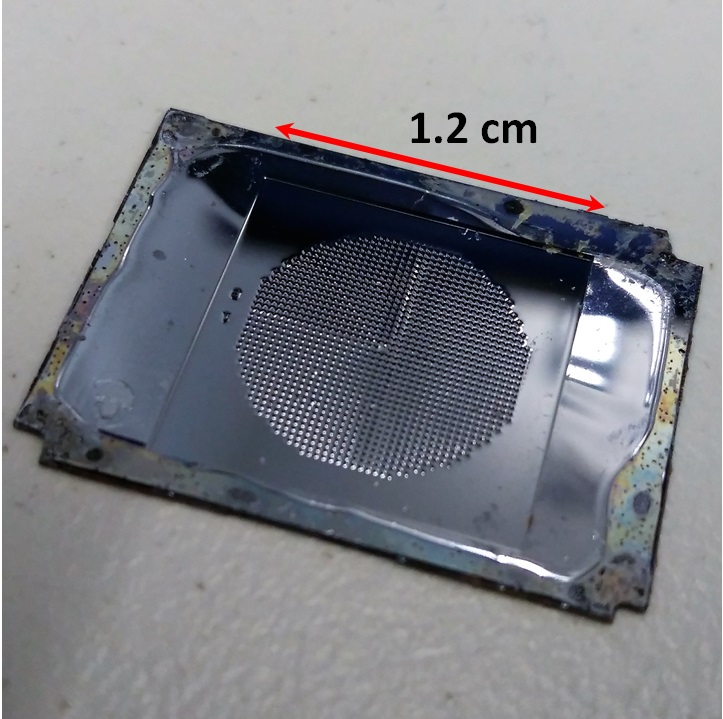}
}
\hspace{0.5cm}
\subfloat[]{
\label{fig:5b}
\includegraphics[height = 3.5cm]{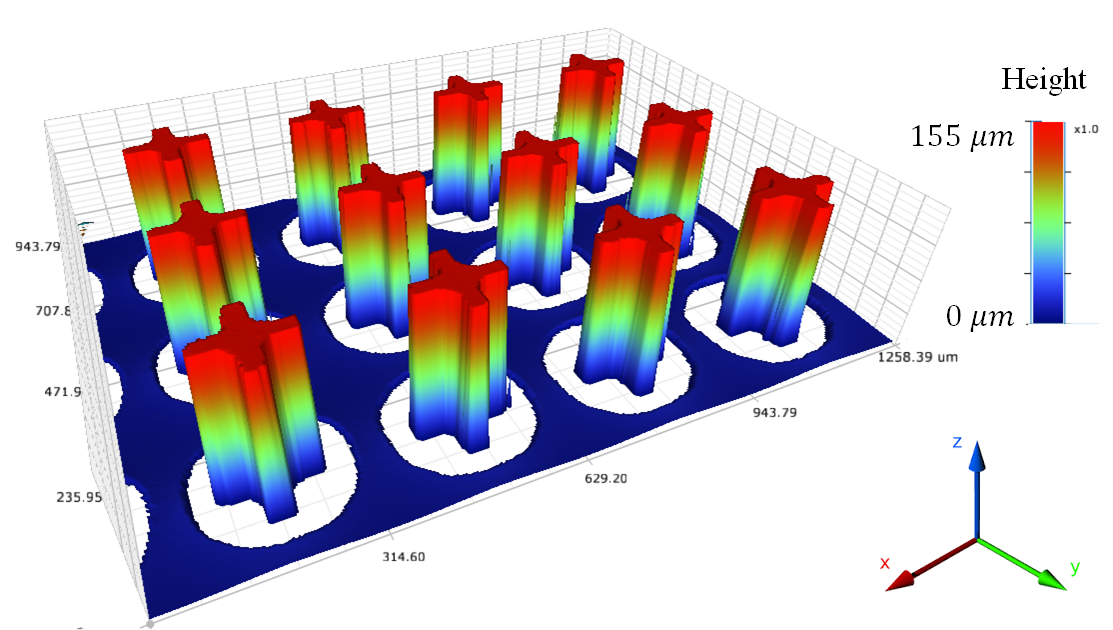}
}
\caption{
\label{fig:5}
(a) Photo of the fabricated metadevice for vortex beam generation.  (b) Depth profile measured using an optical profilometer (Bruker).} 
\end{figure*}

\subsubsection{Fabrication of vortex generator}

A 300~$\mu$m thick intrinsic silicon wafer was used for the fabrication of the optical meta-surface element. The device was fabricated using photolithography followed by reactive ion etching (RIE). The mask for the photolithography was fabricated using the Intelligent micropatterning SF100 XPRESS direct writing system. To achieve a $>10~\mu$m thick resist layer, AZ4562 photoresist was spin coated at 2000~rpm. The large thickness was needed to obtain a high etch depth of 150~$\mu$m by RIE (as the resist acts as a sacrificial mask during this process). The resist was exposed at a dose of 485~$\mu$J/cm$^2$, then developed in AZ726 MIF for 5~minutes. The standard Bosch process was used to etch silicon for the required high aspect ratio pattern. A photograph of the fabricated metadevice and an optical profilometer measurement are shown in Fig.~\ref{fig:5a} and \ref{fig:5b}. The required depth of 150~$\mu$m was achieved after 90~mins of plasma etching.

%__________________Fig. 8
\begin{figure*}[tb]
\centering
\includegraphics[width=0.9\textwidth]{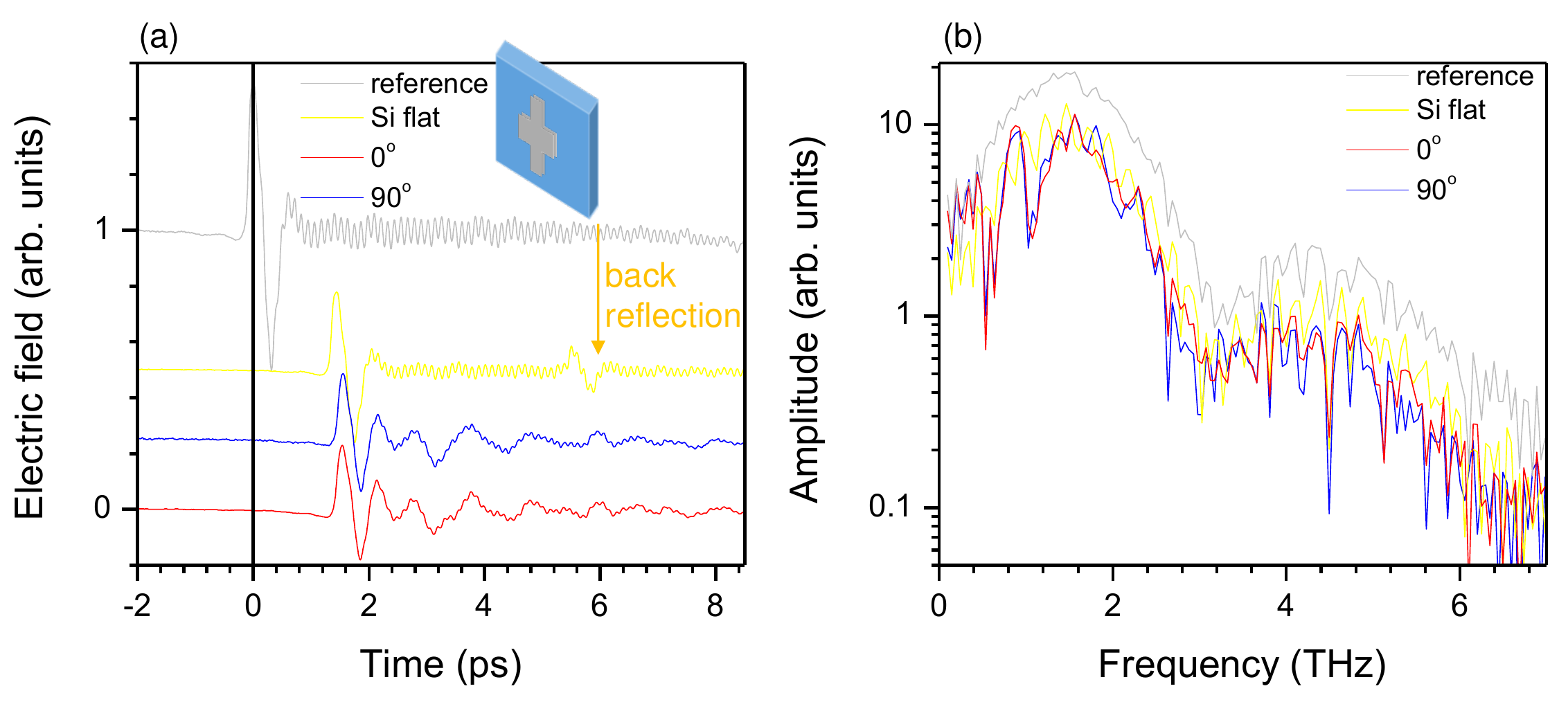}
\caption{Polarization independent action of Si metasurface vortex generator measured by THz time-domain spectroscopy (THz-TDS). (a) Electric field transients (waveforms); reference - is the reflection signal from the GaAs antennae, back-side reflection is shown on the waveform from the flat Si. Two perpendicular polarisations at 0$^\circ$ and 90$^\circ$ shows close  to identical waveforms; a high frequency fringing is the result of Fabri-Perot etalon made of front and back surfaces of Si. (b) Fourier transform of the waveforms shown in (a).}
\label{f-K1}
\end{figure*}

\subsubsection{Characterization by time domain spectroscopy}
 The THz time domain spectroscopy (THz-TDS) was applied to characterise performance of the THz vortex generator shown in Fig.~\ref{fig:5a}.Terahertz transmission spectra of the microstructures were measured using a home-made terahertz time-domain spectrometer (THz-TDS) using a 10-fs Ti:sapphire oscillator with the repetition rate of 80~MHz and output power of 300~mW. The terahertz waves are generated and detected by low-temperature grown GaAs photoconducting antennae
used in a reflection geometry~\cite{Yokota}. The terahertz spectrum spans throughout a broad bandwidth from 0.2 to 15~THz with a dip at 8~THz. The transmittance of the samples are characterized by the ratio of the spectrum of transmitted terahertz wave through the sample with that through the bare silicon substrate with the same thickness. We measured the transmittance at several different positions on the vortex generators. The positions of the incident terahertz waves were confirmed
by the residual of the pump laser reflected from the surface of the generation photoconducting antenna. This guide light was filtered by an additional silicon plate during the measurement. The system was purged with nitrogen to avoid water vapor absorption.
 
%__________________Fig. 9
\begin{figure}[!h]
\centering
\includegraphics[width=0.48\textwidth]{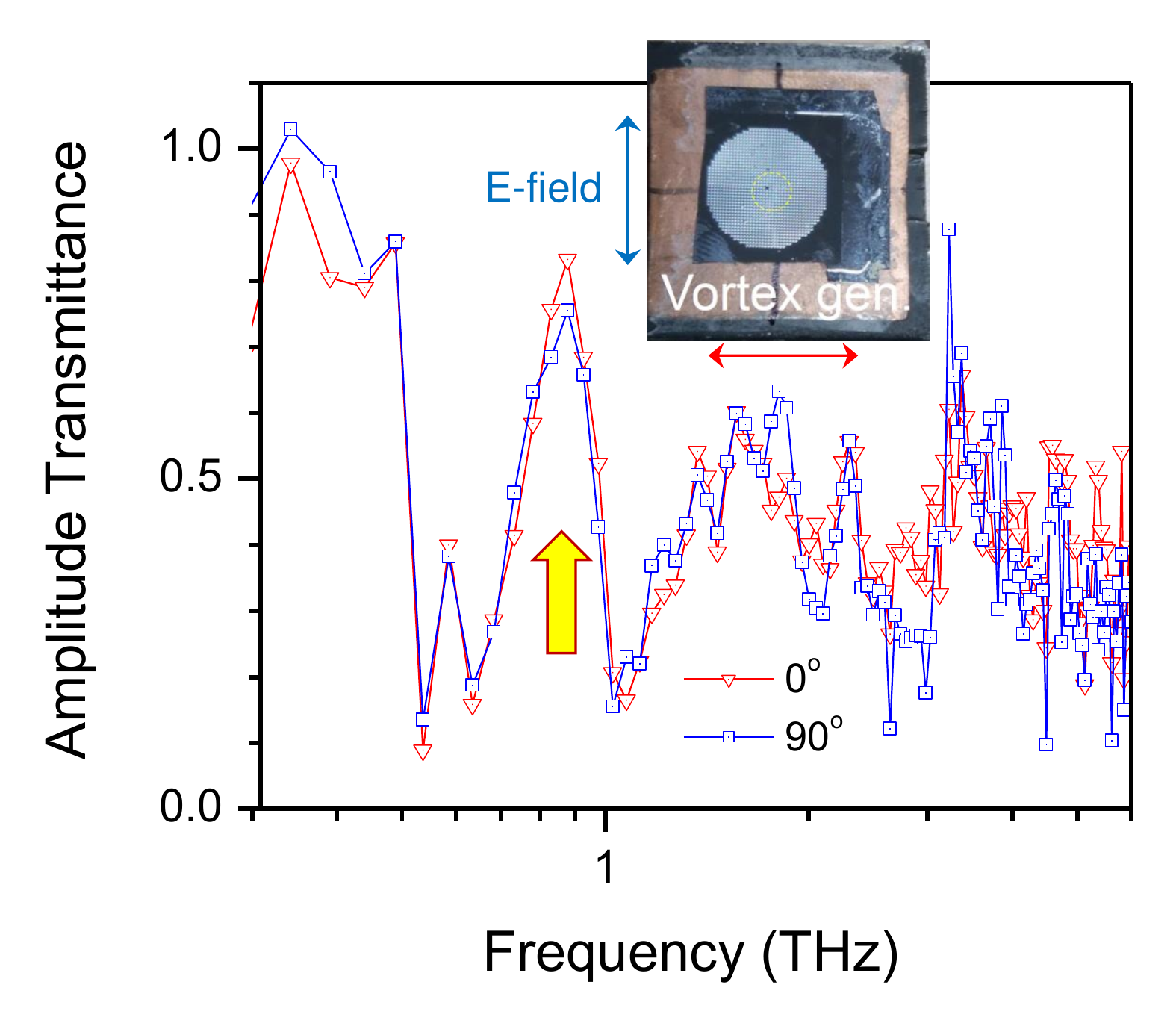}
\caption{
\label{f-K2}
Transmission of the Si metasurface vortex generator. Peak at 0.83~THz (arrow marker) which is close the the designed 0.78~THz. The deviation is due to the slight mismatch between the simulated and fabricated dimensions of the cross resonator.}
\end{figure}

 Almost identical E-field transients were observed in transmission at the two perpendicular polarisations as shown by the waveform and its Fourier transform (spectrum) in Fig.~\ref{f-K1}. Transmission performance of vortex generators designed for maximum $T$ at 0.78~THz are shown in Fig.~\ref{f-K2}. Closely matched spectral performance for two perpendicular polarisations was observed with  transmittance maximum at a slightly shifted 0.8~THz wavelength. The deviation is due to the slight mismatch between the simulated and fabricated dimensions of the cross resonator.

The pattern of cross-antennas is also performing as a spectral filter. A promising application is to combine filter and focusing by flat Fresnel lens pattern to increase signal-to-noise ratio in optical filtering as was recently demonstrated by laser ablation~\cite{Gediminas}.

\section{Conclusions}

Metasurfaces composed of azimutally rotated sections of cross-antennas patterned at different cross lengths $L$ and fixed period $\Lambda$, were used to fabricate optical vortex generators. By using Si, it is demonstrated that high transmission optical elements at the designed wavelength in mid-IR and THz spectral ranges can be made. High refractive index of Si is beneficial for the required overlap of electric and magnetic resonances for the cross-antennas used in this work. The used cross-antenna design is beneficial for polarization insensitive performance of the vortex beam generators. This is a useful virtue for synchrotron applications where the beam has a complex polarization composition~\cite{18jo}.

This study shows that a missing toolbox  of optical elements imparting Orbital Angular Momentum (OAM) onto beams at IR spectral ranges is readily available using a metasurface approach. We used a linear scaling approach to demonstrate the vortex generator at THz wavelength and experimentally demonstrate a high transmission and polarization independent behaviour. Simulation results show desired spatial phase variability needed for beam shaping applications.

\small\section*{Acknowledgements}
The FPA-FTIR imaging experiment was undertaken on the offline FTIR instrument at the Australian Synchrotron IRM Beamline, part of ANSTO, during the approved beamtime for Proposal ID. M13416. This work was performed in part at the Melbourne Centre for Nanofabrication (MCN) in the Victorian Node of the Australian National Fabrication Facility (ANFF). This research was partially funded by JSPS KAKENHI. A.K. is grateful for support via the Russian Science Foundation grant no. 18-79-10091, R.D. thanks IIT Madras--Swinburne joint PhD program. 

\bibliographystyle{elsarticle-num}
\small\bibliography{bib}
\newpage
%\appendix 
% \renewcommand\thefigure{\thesection.\arabic{figure}}    
%\section{Supplement}\label{suppl}
%\setcounter{figure}{0}    
%__________________Fig. suppl 1
%\begin{figure}
%\centering
%\includegraphics[width=0.45\textwidth]{figs/f-time}
%\caption{Time response of \red{displacement $D = \varepsilon E$} of metasurfaces etched in Si for different times shows a comparable performance. Photo insets show actual vortex generators measured.} 
%\label{fig:7}
%\end{figure}

%Figure~\ref{fig:7} shows TDS traces of \red{displacement $D$} from vortex generators made by different etching time from 20 to 40~min.  

\end{document}